\documentclass[preprint,floats,aps,epsfig,nofootinbib,amssymb]{revtex4}

\usepackage{slashed}
\usepackage{graphicx,color}
\usepackage{dcolumn}
\usepackage{bm}
\begin{document}

\title{ Pure Leptonic Gauge Symmetry, Neutrino Masses and Dark Matter}

\author{Wei Chao}
\email{chaow@pku.edu.cn}

\affiliation{$^1$Center for High Energy Physics, Peking University,
Beijing 1000871, China  }

\begin{abstract}
A possible extension of the Standard Model to include lepton number
as local gauge symmetry is investigated. In such a model, anomalies
are canceled by two extra fermions doublet. After leptonic gauge
symmetry spontaneously broken, three active neutrinos may acquire
non-zero Majorana masses through the modified Type-II seesaw
mechanism. Constraints on the model from electro-weak precision
measurements are studied. Due to the $Z_2^{}$ discrete flavor
symmetry, right-handed Majorana neutrinos can serve as cold dark
matter candidate of the Universe. Constraint from dark matter relic
abundance is calculated.
\end{abstract}

\draft

\maketitle

\section{Introduction}

The Standard Model (SM) is in spectacular agreement with all known
experiments. However, it is almost certainly fundamentally
incomplete. The solar \cite{sno}, atmosphere \cite{abcc}, reactor
\cite{kamland} and accelerator \cite{k2k} neutrino experiments have
provided us very convincing evidence that neutrinos are massive and
lepton flavors are mixed. Besides, precisely cosmological
observations have confirmed the existence of non-baryonic cold dark
matter: $\Omega_{\rm D}^{} h^2 =0.1123\pm0.0035$ \cite{darkmatt}.
These two important discoveries can not be accommodated in the SM
without introducing extra ingredients.

A possible extension of the SM is to add three right-handed heavy
Majorana neutrinos ${ N_R^{}}$ so that light neutrino masses can be
generated by the famous canonical seesaw mechanism (i.e., ${
M_\nu^{}= -M_D^{} M_R^{-1} M_D^T}$, where ${ M_D^{}}$ is the Dirac
mass matrix linking left-handed light neutrinos to right-handed
heavy neutrinos and ${ M_R^{}}$ is the mass matrix of ${ N_R^{}}$).
This is the so-called Type-I seesaw mechanism \cite{seesawo}.
Actually, there are three types of tree-level seesaw mechanisms
\cite{seesawo,seesawII,seesawIII} and three types of loop-level
seesaw mechanisms \cite{ernest,masaw,zeesaw,babusaw}. An advantage
of the seesaw mechanism is that they can both explain neutrino
masses and the baryon asymmetry of the Universe with the help of
leptogenesis \cite{leptogenesis}. Putting some discrete flavor
symmetry on the seesaw mechanism, heavy seesaw particles can also
serve as cold dark matter (CDM) candidate
\cite{masaw,radiative,peihonggu}. This builds a bridge between the
dark matter and neutrino physics.

We do not have enough information on the detailed nature of CDM,
except for its relic density. There are major experimental efforts
for direct and indirect detection of dark matter particle beside the
gravitational effect that it has on the Universe, because they must
have some connection to the SM particles. A direct detection probes
the scattering of dark matter off nuclei in the dark matter
detectors, while indirect detection investigate the SM final states
from the annihilation of the dark matter by cosmic ray detectors.

In this paper, we consider the possible extension of the SM to
include lepton number(L) as local gauge symmetry. Two reasons drive
us to investigate this possibility:
\begin{itemize}
\item Baryon number(B) and L are accidental global symmetries in the
SM. B must be broken to explain the baryon asymmetry of the
Universe. L should be broken to accommodate the Majorana masses of
active neutrinos. Investigating the possibility of L as
spontaneously broken local gauge symmetry would help us to study the
origin of small neutrino masses and describe the seesaw scale.
\item Recent results from PAMELA\cite{Pamela,antiproton} and
FERMI\cite{fermionos} suggest there should be pure leptonic
interactions for dark matter to explain the $e^+e^-$ excesses
observed by these experiments. Inspired by Ref.\cite{hemodel}, we
investigate the case dark matter being charged under L, which is
taken as local gauge symmetry.
\end{itemize}
We extend the SM with three right-handed Majorana neutrinos, two
generation fermions doublet as well as pure leptonic gauge symmetry
${ U(1)_X^{}}$ and $Z_2^{}$ discrete flavor symmetry. In this model,
all the quarks have zero $U(1)_X^{}$ charge, while all the leptons
(including right-handed neutrinos) have unit $U(1)_X^{}$ charge.
After $U(1)_X^{}$ gauge symmetry spontaneously broken, right-handed
neutrinos acquire heavy Majorana masses, while three active
neutrinos may acquire non-zero Majorana masses through the modified
Type-II seesaw mechanism. Due to the $Z_2^{}$ symmetry, right-handed
Majorana neutrinos don't couple to the SM fermions, so that they can
be cold dark matter candidate. We study constraints on our model
from neutrino physics and electroweak precision measurements. We
also investigate constraints on the leptonic gauge coupling constant
and the mass of the lightest right-handed neutrino from the dark
matter relic abundance.

The paper is organized as following: Section II is a brief
introduction to the setup of the $U(1)_X^{}$ gauge symmetry. In
section III and section IV, we study constraints on our model from
the neutrino physics and electroweak precision measurements. In
section V, we investigate the possibility of taking the lightest
right-handed neutrino as cold dark matter. Some conclusions are
drawn in section VI.

\section{The Leptonic ${\rm U(1)_X^{}}$ gauge symmetry}
Now we consider the extension of the SM with three right-handed
Majorana neutrinos ${N_R^{}}$, two generation fermions doublet ${
(\Psi_1^{}, \Psi_2^{})_L^{1T}}$, ${ (\Psi_1^{}, \Psi_2^{})_L^{2T}}$
and singlet ${ \Psi_{1R}^{1}, \Psi_{2R}^{1}}$, ${ \Psi_{1R}^{2},
\Psi_{2R}^{2}}$ as well as new gauge symmetry ${ U(1)_X^{}}$ and
$Z_2^{}$ discrete flavor symmetry. To generate Majorana masses for
right-handed neutrinos, one Higgs singlet $\phi$ with ${U(1)_X^{}} $
charge $-2k$ is added to the model. Due to the $Z_2^{}$ discrete
symmetry, right-handed neutrinos do not couple to the SM particles,
such that they can serve as cold dark matter candidate. We also
include one Higgs triplet $\Delta$ with ${ U(1)_X^{}}$ charge $2k$
in our model, so that small but non-zero neutrino masses can be
generated through the modified Type-II seesaw mechanism.
Representations of particles under the symmetry, ${ SU(3)_C^{}
\times SU(2)_L^{} \times U(1)_Y^{} \times U(1)_X^{} \times Z_2^{}}$,
are listed in table I.
\begin{table}[htbp]
\centering
\begin{tabular}{|c|c|c|r|}
\hline Particles & ${ SU(3)_C^{} \times SU(2)_L^{} \times U(1)_Y^{} }$ &${ U(1)_X^{}}$ & $Z_2^{}$\\
\hline ${ (u, d)_L^{}}$ & ${ (~3, ~2, ~{1\over 6})~}$ & $m$ & $1$\\
\hline ${ u_R^{}}$ & ${ (~3,~1,~{2\over 3})}$ & $m$  & $1$\\
\hline ${ d_R^{}}$ & ${ (~3,~1,~-{1\over 3})}$ & $m$ & $1$\\
\hline ${ (\nu, e)_L^{}}$ & ${ (~1, ~2,~ -{1\over 2})~}$ & $k$ & $1$ \\
\hline ${ e_R^{}}$ & ${ (~1,~1,~-{1})}$ & $k$ & $1$ \\
\hline ${ N_R^{}}$ & ${ (~1,~1,~0)}$ & $k$ & $-1$ \\
\hline ${ (\Psi_1, \Psi_2)_L^{1}}$ & ${ (~1, ~2, ~a)~}$ & $b$ & $1$ \\
\hline ${ (\Psi_1^{}, \Psi_2^{})_L^{2}}$ & ${ (~1, ~2, ~-a)~}$ & $b$ & $1$ \\
\hline ${ {\Psi}_{1R}^{1}}$ & ${ (~1,~1,~a+{1\over 2})}$ & $b$ & $1$ \\
\hline ${ {\Psi}_{2R}^{1}}$ & ${ (~1,~1,~a-{1\over 2})}$ & $b$ & $1$ \\
\hline ${ {\Psi}_{1R}^{2}}$ & ${ (~1,~1,~-a+{1\over 2})}$ & $b$ & $1$ \\
\hline ${ {\Psi}_{2R}^{2}}$ & ${ (~1,~1,~-a-{1\over 2})}$ & $b$ & $1$ \\
\hline $H$&$(~1,~2,~1/2)$&$0$&$1$\\
\hline $\Delta$&$(~1,~3,~-1)$&$2k$&$1$\\
\hline $\phi$&$(~1,~,1,~0)$&$-2k$&$1$\\
\hline
\end{tabular}
\caption{ Charges of particle contents in the ${ SU(3)_C^{} \times
SU(2)_L^{} \times U(1)_Y^{} \times U(1)_X^{} \times Z_2^{}}$
scenario. }
\end{table}

Now we investigate how to cancel anomalies of the model. The global
${ SU(2)_L^{}}$ anomaly \cite{globalsu2} requires fermions doublet
to be even. Considering the conditions for the absence of
axial-vector anomaly \cite{avector1,avector2,avector3} in the
presence of ${ U(1)_X^{}}$ and the absence of the
gravitational-gauge anomaly \cite{anog1,anog2,anog3}, which requires
the sum of the ${ U(1)_X^{}}$ charges to vanish, one has
\begin{eqnarray}
&{ SU(3)_C^2 U(1)_X^{}:}& 2m-m-m=0 \ , \\
&{ SU(2)_L^2 U(1)_X^{}:}& {9\over 2}m+{3\over 2}k+ b=0 \ , \\
&{ U(1)_Y^2 U(1)_X^{}~:}&{1\over 2 }m-4m-m+{3\over 2}k-3k+ \sum_i
b\left[2a^2_i-(a_i+{1\over 2})^2 -(a_i-{1\over 2})^2\right]=0 \ ,\\
&{ U(1)_X^2 U(1)_Y^{}~:}&3m^2-6m^2+3m^2-3k^2+3k^2+\sum_i
b^2\left[2a_i-(a_i+{1\over 2})-(a_i-{1\over 2})\right]=0 \ ,\\
&{ U(1)_X^3
~~~~~~~~~:}&3(6m^3-3m^3-3m^3+2k^3-k^3-k^3)+2 (2b^3-b^3-b^3)=0 \ , \\
&{ U(1)_X^{}~~~~~~~~~}:& 3(6m-3m-3m+2k-k-k)+2(2b-b-b)=0 \ , \\
&{ SU(2)_L^2 U(1)_Y^{}:}& {3 \over 2 } - {3 \over 2 } + {1 \over 2 }\sum_i a_i= 0 \; , \\
&{ U(1)_Y^3 ~~~~~~~~~:}& \sum_i^{}\left[2 a^3_i  - ( a_i + {1 \over
2})^3- ( a_i - {1 \over 2} )^3\right]=0 \; , \\
&{ U(1)_Y^{}~~~~~~~~~}:& \sum_i \left[2 a_i  -(a_i + {1\over 2}) -
(a_i -{1 \over 2 })\right]= 0 \; ,
\end{eqnarray}
where $a_i$ is the weak hypercharge of $\Psi^i$ with $a_{1(2)}^{}=
a(-a)$. We find that Eqs. (1), (4)-(6) and (7)-(9) hold
automatically, while Eq. (2) and Eq. (3) are equivalent. As a
result, the upper equations can be simplified to the following
relation:
\begin{eqnarray}
9m+3k+2 b=0 \ .
\end{eqnarray}
Anomalies put no constraint on $a$. We set $a=1/2$ in our paper.
Four interesting scenarios can be derived from Eq. (10):
\begin{itemize}
\item $m=0$ and $k=1$. All the quarks have zero $U(1)_X^{}$ charge.
As a result, ${U(1)_X^{}}$ is a pure lepton number gauge symmetry.
\item $k=0$ and $m=1/3$. All the leptons have zero $U(1)_X^{}$
charge. Such that $U(1)_X^{}$  is a pure baryon number gauge
symmetry.
\item $b=0$ and $k=-3m=-1$. It corresponds to $U(1)_{B-L}^{}$
gauge symmetry, the phenomenology of which has been well-studied.
\item $k=1$, $m=1/3$ and $b = -3$. It correspond to $U(1)_{B+L}^{}$
gauge symmetry.
\end{itemize}

In this paper, we only investigate the phenomenologies of the first
scenario. We will study constraints on the model from neutrino
physics, electroweak precision measurements and cosmological
observations. The phenomenology of the second scenario, which is
interesting but beyond the reach of this paper, will be shown in
somewhere else. The following is the leptonic part of the full
lagrangian
\begin{eqnarray}
{\cal L}_{\rm lep}^{}&=& \overline{ \Psi_L^{} } i \slashed{D}
\Psi_L^{} + \overline{ N_R^{}} i \slashed{D} N_R^{} + \overline{
\Psi_{ R}^{}}i \slashed{D} \Psi_{ R}^{} + \overline{\ell_L^{}} i
\slashed{D} \ell_L^{} + \overline{E_R^{}} i \slashed{D} E_R^{} -
\left[
\overline{ { \over }\ell_L^{}} Y_E^{} H E_R^{}+ \right. \nonumber \\
&& \left.
 \overline{\Psi_L^{i}} Y_{\Psi 1}^{i}\tilde{H} \Psi_{1R}^i +
\overline{\Psi_L^{i}} Y_{\Psi 2}^{i}{H} \Psi_{2R}^i +{1 \over 2}
\overline{N_R^{C}} Y_N^{} \phi N_R^{}+ \overline{\ell^{}_L}
Y_\Delta^{} \Delta \ell_L^{C} +{\rm h.c. } \right] \ ,
\end{eqnarray}
with
\begin{eqnarray}
&&D_{~\mu}^{}= \partial_\mu^{}+ i g_1^{} Y  B_\mu^{} +
ig_2^{}\delta_{{1\over 2} I}^{} \sigma^k  W_\mu^{k} + i g_X^{} Y'
X_\mu^{} \ . \nonumber
\end{eqnarray}
There are no Yukawa interactions between new fermions ($\Psi^{i}$)
and SM particles because of their special ${ U(1)_X^{}}$ quantum
numbers. Then the neutral component of $\Psi$ is stable and can be
dark matter candidate. However, $\Psi$ is strongly coupled to the SM
Higgs boson in our model, i.e., ${\cal O} (Y_\Psi^{})> 1$(of course
it should be smaller than $4\pi$ to satisfy the perturbativity
limit\cite{randall}). Assuming $m_\Psi> 400~ {\rm GeV}$ and
$m_H^{}\sim 200 ~{\rm GeV}$, We can estimate the relic density of
$\Psi$: $\Omega_\Psi^{} h^2 < 2 \times 10^{-3}$, which is quite
small compared with the dark matter relic density, even smaller than
its statistical error. Therefore the contribution of $\Psi$ to the
dark matter relic density is almost ignorable. For simplification,
we will not discuss the phenomenology of $\Psi$ in this paper.

\section{Neutrino masses}

We now investigate the possible origin of Majorana masses for three
active neutrinos in our model. Conventional seesaw mechanisms
explicitly break the lepton number, which is gauged and
spontaneously broken in our model. To overcome this difficulty, we
modify the Type-II seesaw mechanism slightly. The most general gauge
invariant Higgs potential can be written as \footnote{Actually the
full Higgs potential should also contain the following terms:
$\Lambda_1^{} ({ \rm Tr} (\Delta^\dagger \Delta))^2$,
$\Lambda_2^{}{\rm Tr} (\Delta^\dagger\Delta\Delta^\dagger \Delta) $,
$\Lambda_3^{} \phi^\dagger \phi {\rm Tr} (\Delta^\dagger \Delta)$,
$\Lambda_4^{} H^\dagger H{\rm Tr} (\Delta^\dagger \Delta)$ and
$\Lambda_5^{} H^\dagger [\Delta^\dagger, \Delta] H$. Here we assume
their coupling constant are small and thus these terms are
ignorable, jut like what we do with the Higgs potential of the
conventional Type-II seesaw model. The author would like to thank
the anonymous referee for pointing out these terms. }
\begin{eqnarray}
{\cal L}_{\rm Higgs}^{}&=&m_1^2 H^\dagger H+ m_3^2 \phi^\dagger
\phi+M^2_\Delta {\rm Tr}\left(\Delta^\dagger \Delta\right) +{1\over
2}\lambda_1^{}(H^\dagger H)^2 +  {1\over 2}\lambda_3^{}(\phi^\dagger
\phi)^2 \nonumber \\ && + \lambda_5^{} (\phi^\dagger \phi)(H^\dagger
H) +[\lambda_7^{} \phi H^T i\sigma_2^{} \Delta H +{\rm h.c.}] \; .
\end{eqnarray}
Here $H$ plays the role of the SM Higgs doublet. On the contrary to
the conventional Type-II seesaw mechanism, the last term in Eq. (12)
conserves the lepton number. When $\phi$ gets vacuum expectation
value(VEV), ~$U(1)_X^{}$ gauge symmetry is broken down and
right-handed neutrinos acquire Majorana masses. We set $\langle H
\rangle =v$ and  $\langle \phi \rangle =v'$. After imposing the
conditions of global minimum, one obtains
\begin{eqnarray}
\langle H \rangle^2\approx{\lambda_3^{} m_1^2 -\lambda_5^{} m_3^2
\over \lambda_5^2 - \lambda_1^{} \lambda_3^{}}\ , \hspace{1cm}
\langle \phi\rangle^2  \approx{\lambda_1^{} m_3^2 -\lambda_5^{}
m_1^2 \over \lambda_5^{2}-\lambda_1^{} \lambda_3^{}} \ ,
\hspace{1cm} \langle \Delta \rangle^{}=-{\lambda_7^{} v' v^2 \over 2
M_\Delta^2} \ .
\end{eqnarray}

The light neutrino mass matrix is then
\begin{eqnarray}
M_\nu^{} = Y_\Delta^{} \langle \Delta \rangle =- Y_\Delta^{}
{\lambda_7^{} v' v^2 \over 2 M_\Delta^2}\; .
\end{eqnarray}
Present constraint on the neutrino mass matrix from neutrino
oscillation experiments and cosmological observations is ${\cal O}
(M_\nu^{}) \sim 0.1 {\rm eV}$\cite{pdg}. By setting $\langle \phi
\rangle = 1~ {\rm TeV}$ and $M_\Delta^{} = 10^{8}~ {\rm GeV}$, one
has $Y_\Delta^{} \lambda_7^{} \sim 10^{-2}$.

\section{Electroweak precision Measurement constraints}

We now perform an analysis of the electroweak precision observable
on our model. The most stringent restrictions come from the $S$ and
$T$ parameters \cite{s and t}, which can be expressed by the
following equations
\begin{eqnarray}
S&\approx&\sum_{\alpha=1}^2 { 1 \over 6\pi } \left[ 1-Y_\alpha^{}
\ln \left({M_{\Psi1}^\alpha \over M_{\Psi2}^\alpha }\right)^2
\right]\
,\\
T&\approx&\sum_{\alpha=1}^2 {1 \over 16 \pi s^2 c^2 M_Z^2 } \left[
(M_{\Psi1}^\alpha)^2 + (M_{\Psi2}^\alpha)^2  - {2
(M_{\Psi1}^\alpha)^2 (M_{\Psi2}^\alpha)^2 \over (M_{\Psi1}^\alpha)^2
- (M_{\Psi1}^\alpha)^2 } \ln \left( { M_{\Psi1}^\alpha \over
M_{\Psi2}^\alpha} \right)^2 \right] \ ,
\end{eqnarray}
where $s= \sin \theta_W^{}, c= \cos \theta_W^{} $ with $\theta_W^{}$
the Weinberg angle. $Y_\alpha^{} $ is the weak hypercharge of
$\Psi_\alpha^{}$ with $Y_{1, 2}=-1/2, 1/2$. To obtain a better
understanding of the importance of the $S$ and $T$ parameters in
constraining our model, we show in Fig. 1 the $90 \%$ Confidence
Level contour (ellipse) in the $(S, ~T)$ plane, as obtained from the
Electroweak Working Group \cite{pdg}, together with the new heavy
fermion's predictions. In plotting the figure, we have assumed the
masses of heavy fermions lie  in the range [400, 1000] GeV. It is
clear from the figure that there is sizeable region in the parameter
space lying within $90\%$ Confidence Level contour.

\begin{figure}[h]
\begin{center}
\includegraphics[width=11cm,height=8cm,angle=0]{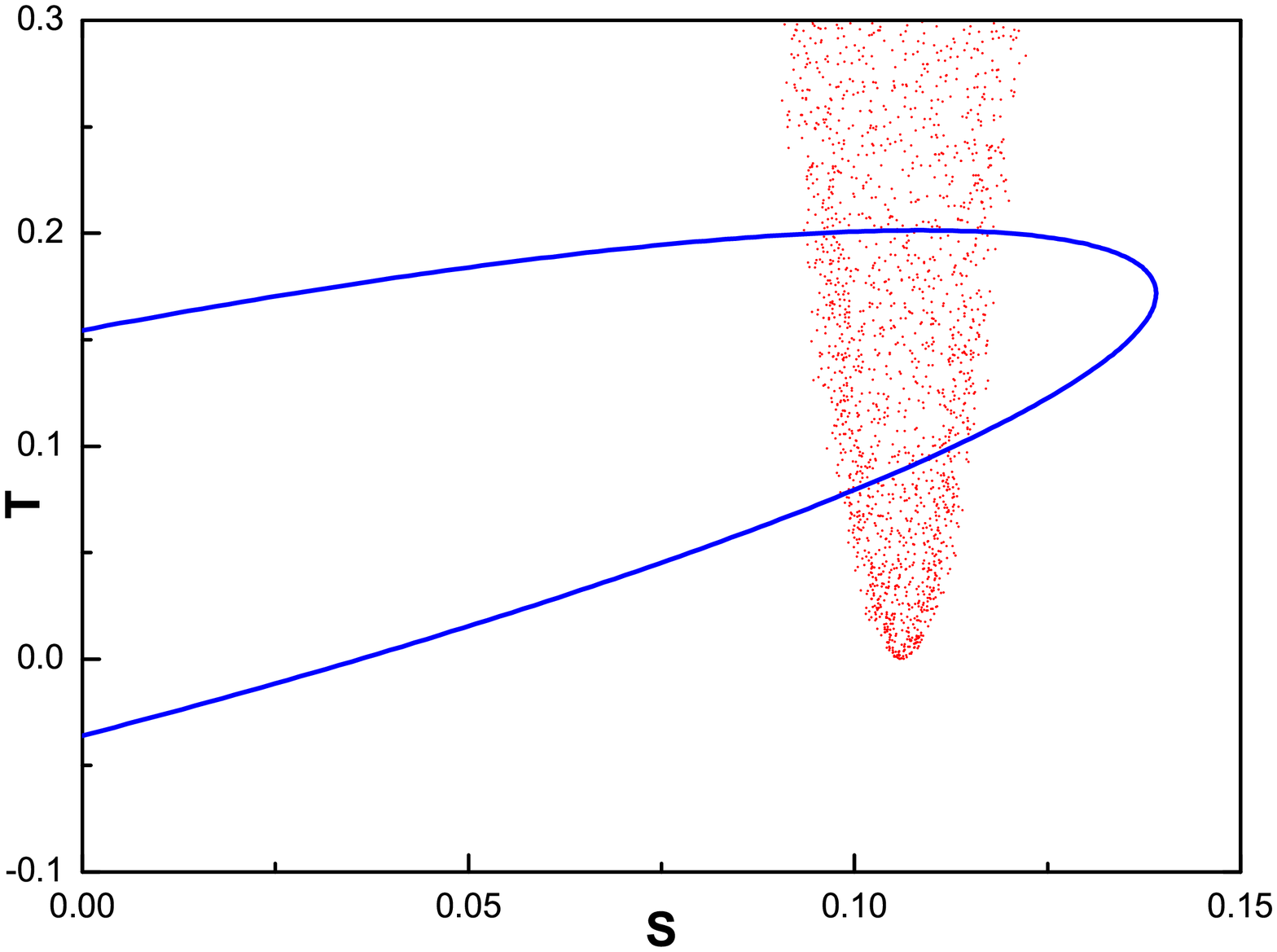}
\end{center}
\caption{Numerical illustration of the $90\%$ Confidence Level
contour (ellipse) and new heavy fermions' predictions in the $(S,
T)$ plane. The dots indicate points where all the heavy fermions lie
between $400$ GeV and $1000$ GeV.}
\end{figure}

We proceed to consider the constraint on the model from the muon
g-2. There has been a long history in measuring and calculating the
muon abnormal magnetic moment $a_\mu^{}$. In particular the steadily
improving precision of both the measurements and the predictions of
$a_\mu$ and the disagreement observed between the two have made the
study of $a_\mu$ one of the most active research fields in particle
physics in recent years. The final result of the ``Muon g-2
Experiment"(E821) for $a_\mu^{}$ reads  $a_\mu^{\rm
exp}=(11659208\pm6)\times10^{-10} \  $ \cite{g-2}, which deviates
from the SM prediction by
\begin{eqnarray}
\Delta a_\mu^{}=a_\mu^{\rm exp}-a_\mu^{\rm SM}=22(10) \times
10^{-10}\ .
\end{eqnarray}
In our model, heavy neutral gauge boson $Z'$ will contribute to
$\Delta a_\mu^{}$, which is
\begin{eqnarray}
\Delta a_\mu^{}= {g_X^2 \over 8\pi^2 } \int_0^1  dx{ 2m_\mu^2
x^2(1-x) \over x^2 m_\mu^2 + (1-x) M_{Z'}^2}  \ ,
\end{eqnarray}
where $M_{Z'}^{}$ is the mass of $Z'$. There are a lot of works on
the heavy neutral gauge bosons. For a recent review, see
Ref.~\cite{lankaer}, and recent studies $Z'$ at Tevatron and LHC
\cite{zLHC}. In our model, the mass of the new gauge boson $Z'$ is
given by $M_{Z'}^{}=2 g_X^{} v'$. To satisfy the experimental lower
bound, $M_{Z'}^{}/g_X^{} > 5 \sim 10 ~{\rm TeV}$ \cite{consz}. Given
$M_{Z'}^{}>500 {\rm GeV}$ and ${\rm g_X^{}<0.1}$, we can find the
biggest contribution to $\Delta a_\mu^{}$ is $4\times 10^{-12}$,
which is far below the discrepancy listed in Eq. (17). In short
there is almost no strong constraint on the $g_X^{}$ and $M_{Z'}^{}$
from the muon g-2.

\section{dark matter}
In our model, the lightest heavy Majorana neutrino serves as dark
matter candidate. Assuming $M_N^{} < M_\Psi^{}$, there are two
dominant annihilation channels: $\overline NN\rightarrow Z'
\rightarrow \overline \ell \ell$ and $\overline NN\rightarrow
\varphi \rightarrow \bar q q (\bar \ell \ell)$, where $\varphi$ is
the SM Higgs. The second channel is heavily suppressed by the mixing
between $\phi$ and $H$, such that we only consider the first
channel. Ignoring charged lepton masses, one can write down the
thermal average of the interaction rate $\sigma v$ in
non-relativistic limit
\begin{eqnarray}
\langle \sigma v_{\rm M{\o}l} \rangle = \langle \sigma v_{\rm
lab}^{} \rangle \approx a^{(0)} + {3 \over 2} a^{(1)}  x^{-1}_f= {
m^2 g_X^4 \over 4 \pi \left[( 4m^2 - M_{Z'}^2) + M_{Z'}^2
\Gamma_{Z'}^2 \right]} x^{-1}_f \; ,
\end{eqnarray}
where $x_f^{}=M_N^{}/T_f^{}$ with $T_f^{}$ the freeze-out
temperature of the relic particle. In assumption $M_{Z'}^{} < 2
M_\phi^{}$, the decay width of $Z'$ is
\begin{eqnarray}
F_{Z'}^{} = { g_X^2 M_{Z'}^{} \over 8 \pi} \; .
\end{eqnarray}
The present density of dark matter is simply given by
$\rho_N^{}=M_N^{} s_0^{} Y_\infty^{}$, where $s_0^{} = 2889.2~ {\rm
cm}^{-3}$ is the present entropy density and $Y_\infty^{}$ is the
asymptotic value of the ratio $n_N^{}/ s_0 $ with $ Y_\infty^{-1} =
0.264 \sqrt{g_\ast} M_{Pl}^{} M_N^{} (a^{(0)}+3a^{(1)}/4x_f)$, where
$g_*^{}$ accounts the number of relativistic degrees of freedom at
the freeze-out temperature. The relic density can finally be
expressed in terms of the critical density
\begin{eqnarray}
\Omega h^2 \simeq { 1. 07 \times 10^9~ {\rm GeV^{-1}} \over
M_{Pl}^{}} {x_f \over \sqrt{g_\ast}} {1 \over a^{(0)}+3 a^{(1)}/4x_f
} \; ,
\end{eqnarray}
where $h$ is the Hubble constant in units of $100 ~{\rm km} / {\rm
s\cdot Mpc}$ and $M_{Pl}^{}= 1.22 \times 10^{19} ~{\rm GeV}$ is the
Planck mass. The freeze-out temperature $x_f^{}$ can be estimated
through the iterative solution of the following equation
\cite{darkrep}
\begin{eqnarray}
x_f^{} = \ln \left[ c(c+2) \sqrt{45\over 8} {g \over 2\pi^3} {{
M_{Pl}^{}M_N^{}} \langle \sigma_{ann}v_{rel}\rangle \over
\sqrt{g_\ast x_f^{}} } \right]\simeq {\rm ln}{0.038 M_{Pl}M_N^{}
(a^{(0)}+3a^{(1)}/2x_f) \over \sqrt{g_\ast x_f}}
\end{eqnarray}
where $c$ is the constant of order one determined by matching the
late-time and early-time solutions.
\begin{figure}[h]
\begin{center}
\includegraphics[width=10cm,height=7cm,angle=0]{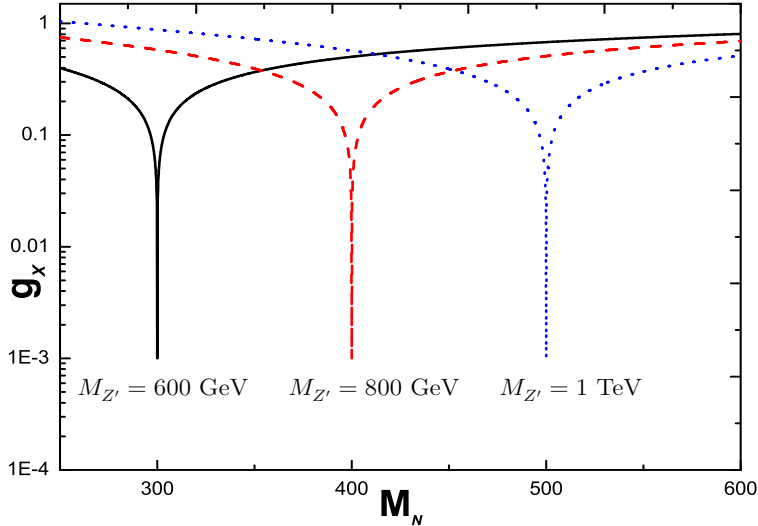}
\end{center}
\caption{$g_X^{}$ versus dark matter mass $m_N^{}$ constrained by
the dark matter relic abundance.} \label{relic}
\end{figure}

We set $x_f^{}$ equals to $20$, a typical value at freeze-out for
weakly interacting particles. In FIG. ~\ref{relic}, we plot $g_X^{}$
versus the mass of the lightest right-handed neutrino, $M_N^{}$,
constrained by the dark matter relic abundance. The solid, dashed
and dotted lines correspond to $M_{Z'}^{}= 600 ~{\rm GeV}$, $800
~{\rm GeV}$ and $1000 ~{\rm GeV}$, separately. One finds poles at $
M_N^{} = 1/2 M_{Z'}^{}$, where the annihilation cross section is
resonantly enhanced. All the  experimental constraints can be
fulfilled near these region.

Notice that heavy Majorana neutrinos only annihilate into leptons,
our model could explain $e^+ , e^-$ excess reported by PAMELA
\cite{Pamela,antiproton} and Fermi \cite{fermionos} with resonant
enhancement \cite{zuoweiliu,breitwig,wanlei} as boost factor. For
similar analysis on this subject, see Ref \cite{fox, xiaojun, made}
for detail.

\section{conclusions}
We have investigated the possibility of taking the lepton number as
local gauge symmetry. In such a model, at least two fermions doublet
are needed to cancel the anomaly. We have introduced a modified
Type-II seesaw mechanism to generate Majorana masses for three
active neutrinos. Constraints from electroweak precision
measurements were studied. The result shows that there are adequate
parameter space for our model. Taking heavy Majorana neutrinos as
dark matter candidate, we have studied the constraint on the
leptonic gauge coupling constant and the mass of the lightest
right-handed neutrino from dark matter relic abundance.

\begin{acknowledgments}
The author thanks to S. Luo and W. L. Guo for helpful discussion
This work was supported in part by the National Natural Science
Foundation of China.
\end{acknowledgments}
\section*{Note added}
During the completion of this work, Ref.~\cite{mbwise}, which
investigate the $U(1)_B^{} \times U(1)_L^{}$ gauge symmetry,
appeared. We built similar frameworks, but focused on different
phenomenologies.


\end{document}